\documentstyle[prl,aps,twocolumn,graphicx]{revtex}

\begin{document}

\title{Multi-mode description  of an interacting Bose-Einstein condensate} 

\author{Krzysztof G{\'o}ral$^{1,3}$, Mariusz Gajda$^{2,3}$, and 
Kazimierz Rz{\c a}\.zewski$^{1,3}$}
\address{$^1$ Center for Theoretical Physics, $^2$ Institute of
Physics, $^3$ College of Science,\\ Polish Academy of Sciences, Aleja
Lotnik\'ow 32/46, 02-668 Warsaw, Poland}


\maketitle

\begin{abstract}
We study the equilibrium dynamics of a weakly interacting Bose-Einstein 
condensate trapped in a box. In our approach we use a semiclassical
approximation
similar to the description of a multi-mode laser. In dynamical equations
derived from a full $N$-body quantum Hamiltonian we substitute all creation
(and annihilation) operators (of a particle in a given box state) by
appropriate c-number amplitudes.  The set of nonlinear equations obtained
in this way is solved numerically. We show that on the time scale of a few
miliseconds the system exhibits relaxation -- reaches an equilibrium with 
populations of different eigenstates fluctuating around their mean values.
\end{abstract}



Bose-Einstein condensation of an ideal gas is a famous example of a
phase
transition which relies purely on quantum statistics.  The recent
experimental achievement of Bose-Einstein condensation in a dilute gas of
trapped alkali atoms \cite{BEC_exp_Rb,BEC_exp_Na,BEC_exp_Li,BEC_exp_H}
triggered a revival of interest in various
properties of the condensate. Some old but fundamental problems of
statistical description of the Bose-Einstein condensate have been extensively
studied. In particular, the issue of microcanonical and canonical 
fluctuations of a noninteracting condensate was solved successfully
\cite{Navez,Gajda,GrossmannPRL,GrossmannOpt,Politzer,WilkensJMO,WilkensOpt}. On
the
contrary, a closed system
description of
fluctuations of the weakly interacting condensate seems to be still a
controversial problem at least at temperatures close to the critical one
\cite{Stringari,Idziaszek,Zwerger,Scully}.  The standard mean field
description based on the Bogolubov approach fails \cite{StringariRMP} at
the
critical temperature, as then the order parameter vanishes.  On the other hand
many excited states become highly occupied particularly in the case of
realistic
systems consisting of a finite number of atoms.  The complexity of the
problem
grows further when one is interested in an estimation of the
characteristic time scale of the condensate
fluctuations\cite{GrahamPRA,GrahamPRL}. 
The calculation of the two-time correlation function requires
obviously a dynamical approach.

In the following, we present an approach that allows the study of the time
evolution of an interacting Bose gas.  The problem is very difficult
and existing approaches lead to numerical algorithms which are hard to
implement. The first one \cite{Gardiner,Jaksch} invokes quantum kinetic
theory and leads to a quantum Boltzmann master equation.  The second one
\cite{Holland,Walser} is based on the separation of the ``classical" mean
field
describing the condensate from its fluctuations. The fluctuations are
quantal in nature and the dynamics couples averages of both their normal
and anomalous parts to the condensate mean field. None of the two
mentioned methods seems to be capable of handling a realistic case of a
large number of particles at a  relatively high temperature.  

The first problem one encounters in describing the interacting
condensate is  the real N-body nature of the system. The link between this
problem
and the traditional ``giant matter wave" description is not
straightforward.  The
only exception is the case of a gas trapped in a box with periodic
boundary
conditions. Here the symmetry of the potential uniquely enforces the form
of the eigenstates of the one-particle density matrix regardless the
interaction. Due to the symmetry, the condensate can be associated
with the ground state of the trap. Therefore, to avoid all
ambiguities related to the identification of the condensate we are
going to study here the system in a box with periodic boundary conditions 
\cite{comment}.

We want to present our approach rooted in the theory of  multi-mode
lasers.  Let us start with a general formulation of the N-body
problem.  The second-quantized Hamiltonian for the atomic system confined in a
box with periodic boundary conditions and interacting via pairwise
contact potential may be written in the following form:
\begin{equation}
\label{H1}
H =\int {\rm d^3} r \, \Phi^{\dagger}\frac{p^2}{2m}\Phi
+  \frac{V \hbar g}{2} \int {\rm d^3} r \; \Phi^{\dagger}
\Phi^{\dagger} \Phi \Phi,
\end{equation}
where $\Phi$ is an atomic field operator, $V=L^3$ is a volume of the
system ($L$
being a size of the box) and $g=\frac{4\pi\hbar a}{m V}$ characterizes the
atom-atom interactions in the low-energy, s-wave approximation ($a$ being the
scattering length and $m$ -- the mass of the atom).  The field $\Phi$ is expanded in
natural modes of the system -- the plane waves:
\begin{equation}
\label{expan}
\Phi({\bf r})=\frac{1}{\sqrt{V}}\sum_{{\bf k}}\exp(-i{\bf k}
\cdot{\bf r})a_{{\bf k}} \ ,
\end{equation}
where $a_{{\bf k}}$ are bosonic annihilation operators, and ${\bf
k}=\frac{2\pi}{L}{\bf n}$ with $n_{i}=0,\pm 1,\pm 2,\ldots$ ($i=x,y,z$). With
this substitution the Hamiltonian assumes its final form:
\begin{equation}
\frac{H}{\hbar} = \xi \sum_{{\bf k}}n^2 \, 
a^{\dagger}_{{\bf k}} a_{{\bf k}} 
+\frac{1}{2}g \sum_{{\bf k},{\bf k'},{\bf k''}} a^{\dagger}_{{\bf k}+{\bf
k'}-{\bf k''}} a^{\dagger}_{{\bf k''}}a_{{\bf k'}}a_{{\bf k}},
\end{equation}
where $\xi=\frac{\hbar}{2m}(\frac{2\pi}{L})^2$. After elimination of a fast
time dependence with the substitution ${a}_{{\bf k}}=\exp(-i\xi n^2 t)
{\alpha}_{{\bf k}}$, the Heisenberg equations of motion for the operators
$\alpha_{{\bf k}}$ acquire the following form:
\begin{equation}
\label{Heisenberg}
\dot{\alpha}_{\bf k} = -  ig \sum_{{\bf k'},{\bf
k''}} \exp\left[2 i\xi ({\bf n}-{\bf n'})({\bf n}-{\bf n''})
t\right] \alpha^{\dagger}_{{\bf k'}+{\bf k''}-{\bf k}}\alpha_{{\bf
k'}}\alpha_{{\bf k''}}. 
\end{equation}

Solving the nonlinear operator equations
(\ref{Heisenberg}) is difficult.  The complexity of the problem obviously
requires some approximation. A semiclassical approximation is particularly
well suited for the
description of a finite system at temperatures below the critical one,
except the region close to the absolute zero.

The semiclassical approximation consists in replacing all operators
$\alpha_{\bf k}$ by c-number complex amplitudes (we are not going to introduce
a separate notation for corresponding complex fields). At
very low temperatures only the lowest lying states are macroscopically
occupied
and quantum fluctuations in excited states become important (see 
\cite{Holland} for comparison). Therefore, the relevant range of
temperatures for the applicability of our
model excludes temperatures close to zero. From the
viewpoint of the Bogolubov method \cite{Fetter}, such an approach is legitimate
as indeed many modes are macroscopically populated, i.e. their occupation
is greater than quantum fluctuations. 

The semiclassical approximation leads to nonlinear differential equations which
must be solved numerically.  The first observation is that 
Eqs.(\ref{Heisenberg}) can be viewed as a set of Hamilton equations for
the
complex degrees of freedom. Our approximate dynamics  preserves the
number of particles: $N = \sum_{\bf k} \alpha_{\bf k}^* \alpha_{\bf k}$, as
well as the total energy of the system. It therefore corresponds to a genuine
microcanonical description. Moreover, the resulting equations resemble the
famous Fermi-Pasta-Ulam \cite{FPU} problem of a system of harmonic
oscillators
coupled by a nonlinear interaction. A one-dimensional version of this dynamics
has been studied recently \cite{Villain} in the context of the pure
Bose-Einstein condensate ($T=0$) reloaded from a harmonic into a rectangular
trap. Equations studied here, however, in spite of a formal analogy, 
describe quite a different physical situation. Our complex amplitudes are
not expansion coefficients of the condensate wave function in some
convenient basis. They
represent a number of coupled ``mean fields" -- a natural extension of the
condensate mean field of the Bogolubov approach. Let us notice that if we start
with $100\%$ occupation of the ${\bf k}=0$ mode and keep only this mode
in the model we simply recover the standard Gross-Pitaevskii equation for
the interacting condensate.

For the initial conditions we assume ``Bose-Einstein-like" occupation of
different trap states. While calculating the energy in order to determine   
this distribution, however, we
neglect the energy of interparticle interactions. This is evidently not the
equilibrium distribution for the interacting system. Moreover, a population of
individual states does not specify initial conditions uniquely. It defines
the
modulus of the corresponding amplitude but says nothing about its
phase. In our
approach each mode is assigned an initial, randomly chosen, phase. Any
subsequent dynamics depends on the initial phases and, in a sense, a single
simulation describes a single experimental realization. Microcanonical
expectation values  would require, therefore, an average over these 
initial phases. As we have checked, instead of doing this, it is enough to
start with
some random phases and trace the system evolution for a sufficiently long time.
The observed self-averaging can be attributed to the ergodicity of the studied
dynamics. Contrary to the 1-D analogue which is completely integrable
\cite{Chirikov}, the 3-D version of the dynamics may be chaotic. We make
use of this fact in our simulations and avoid averaging over phases of
the complex amplitudes.

\begin{figure}[hbp]
\begin{center}
\includegraphics[width=\columnwidth,clip]{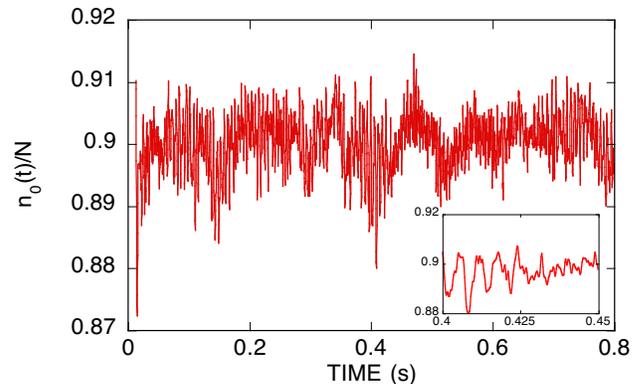}
\caption{Condensate occupation as a function of time for total energy
per particle $E/\hbar=539 Hz$.}
\label{fig1}
\end{center}
\end{figure}

Values of the parameters in the model are $\xi=$71.373 Hz, $N=10^5$ and
$g=$0.018 Hz (the atomic mass and the scattering length are those of $^{87}$Rb
and the size of the box is equal to the Thomas-Fermi radius of a condensate of
$N$ atoms in a trap with frequency of $\omega_{0}=2\pi \; 80$ Hz). We
performed
our calculations for the model with 729 modes ($n_{i}= -4,\ldots,4$,
$i=x,y,z$). Further increasing of the number of modes does not lead to a
substantial change in the results for the case studied in this paper.  Our
calculations show that after a time of the order of a few miliseconds the
system reaches a dynamical equilibrium. The mean occupation of the
condensate ({\bf k}=0 mode) stabilizes at some value and on larger time
scales (of the order of a second) it only fluctuates around
this mean value -- see Figure \ref{fig1}. 

The similarity to the Fermi-Pasta-Ulam problem may cast some doubts on the
genuine ergodicity of the dynamics investigated in this paper
. Originally, Fermi,
Pasta, and Ulam intended to test the ergodic hypothesis in the chain of coupled
harmonic oscillators. However, they observed a quasiperiodic behavior
identified by the returns of energy to the initial (lowest) mode. This
kind of
behavior has been discovered also by J.H. Eberly and
co-workers \cite{Eberly} in a
resonantly coupled system composed of a two-level atom and a single mode
of a monochromatic electromagnetic field being initially in a coherent
state.
Occurrence of the so called {\it revivals} in the system proves the
quantum
nature of the electromagnetic field.  In our calculations the largest time
scale for which we have studied the dynamics was of the order of one
second.  On this time scale we did not observe any revivals but this,
obviously, does not exclude the possibility of revivals on much larger
time scales.
In fact our numerical simulations involve a finite number of modes and so 
the numerical implementation inevitably leads to a quasiperiodic
evolution.

Since the evolution (see Eqs.(\ref{Heisenberg})) is Hamiltonian, the most
natural choice of independent variables are the energy and the particle
number.
The number of particles was fixed through
all our calculations ($N=10^5$) and the energy of the system was the control
parameter. Traditionally, however, the temperature, not the energy, is
used as
an independent thermodynamic variable. Calculation of the microcanonical
temperature requires monitoring of the entropy of the system for different
energies.  Although, in principle, this can be done \cite{Davis}, we
think that the energy is a much better characteristic of the system. The
reason is
twofold. First of all, it is not obvious whether the ergodic hypothesis
can be
applied in the studied case and therefore whether the notion of the
temperature
can be unambiguously introduced. Secondly, in realistic experiments it
is
rather the final energy of thermal atoms which is detected in 
destructive time-of-flight measurements. The temperature is a parameter
fitted
to the observed velocity distribution. 

\begin{figure}[hbp]
\begin{center}
\includegraphics[width=\columnwidth,clip]{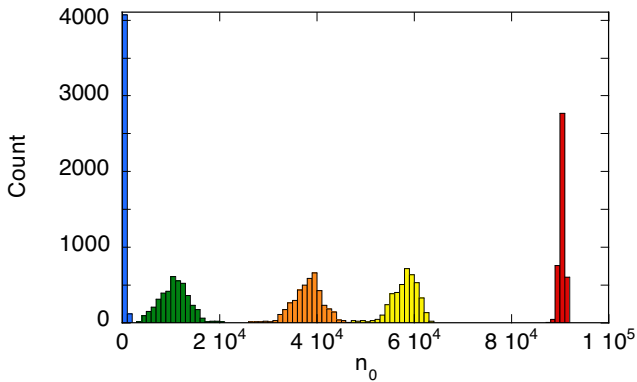}
\caption{Probability distribution of the condensate (${\bf k}=0$
mode) population. Different colors signify different total energies per
particle: blue $E/\hbar=2036$ Hz, green $E/\hbar=1680$ Hz,
orange $E/\hbar=1424$ Hz, yellow $E/\hbar=1164$ Hz, red $E/\hbar=539$ Hz.}
\label{fig2}
\end{center}
\end{figure}

In Figure \ref{fig2} we present probability distributions of the 
condensate population. 
The different colors signify different total energies. Let us stress two
main features which can be seen in Figure \ref{fig2}: (i) positions of the
maxima of these distributions correspond to a mean ({\it
time-averaged}) occupation of the Bose-Einstein condensate, (ii) the
spread of these distributions around the most probable value is a measure
of condensate fluctuations. The figure clearly indicates that
the condensate population decreases monotonically with energy while
fluctuations become larger.

\begin{figure}[hbp] 
\begin{center}
\includegraphics[width=\columnwidth,clip]{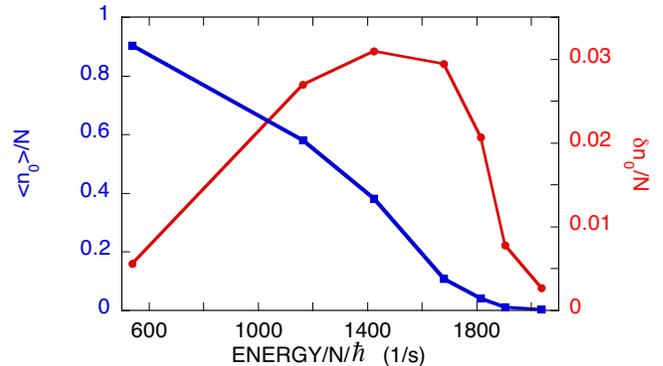}
\caption{Condensate occupation (in blue) and fluctuations (in red) versus
total energy per particle.}
\label{fig3}
\end{center}
\end{figure}

The occupation and the fluctuations of the interacting condensate are
shown in detail in Figure \ref{fig3}, where the mean occupation of the
condensate is depicted (in blue). We see that the condensate disappears at
the energy per particle close to $E/\hbar=2000$ Hz.


Figure \ref{fig3} also presents fluctuations of the condensate (in
red). Both the mean occupation and the fluctuations of the condensate are
smooth functions of the
energy. They do not show any discontinuity signifying a phase transition
because they correspond to a finite system of $N=10^5$ atoms. Let us
notice,
however, that fluctuations reach the maximum value at the energy per
particle close to
$E/\hbar=1450 Hz$. Moreover, this value of energy corresponds to the
inflection point
of the mean occupation of the condensate. Both curves
distinguish the same characteristic value of the energy. Close to this energy
the system undergoes rapid changes.
This characteristic energy corresponds to the critical energy for the
Bose-Einstein condensation.

In conclusions: first of all, we have proposed a very efficient approach
to the time evolution  of an interacting Bose-Einstein condensate. This
approach is valid for a wide range of temperatures except for very small
ones. We have shown that the method works very well in a realistic and
relevant range of parameters: energy, number of particles, and interaction
strength. We calculated the mean occupation as well as the fluctuations of
the interacting condensate by averaging the corresponding time-dependent
quantities. We did not explore all possible applications of our 
method. Our aim was rather to show its potential. We believe that  the
semi-classical approach presented here is perfectly suited for studying
many properties of an interacting condensate which so far, due to
their  complexity, were beyond direct theoretical investigations. In
particular, it may shed new light at the unresolved problem of nucleation
of a
condensate and other dynamical phenomena. The method is easily extendible
to more than one field which allows to study coupled (via
photoassociation or Feshbach resonances) atomic and molecular systems (see
\cite{GoralMolecules}) or other mixtures of Bose gases at finite
temperatures.

K.R. and K.G. are supported by the subsidy of the
Foundation for Polish Science. M.G.  acknowledges support by Polish KBN grant
no 2 P03B 078 19. Part of the results has been obtained using computers at the
Interdisciplinary Centre for Mathematical and Computational Modeling (ICM) at
Warsaw University.


\begin{thebibliography}{}
\bibitem{comment} The case of boundary conditions different from the
periodic ones (e.g. a rectangular trap) presents an interesting and
challenging problem. In this case there are no universal eigenstates of a
one-particle density matrix and therefore the definition of a condensate
is unclear.
\end{thebibliography}
\end{document}